\documentclass[
reprint,
superscriptaddress,
altaffilletter,
nofootinbib,
amsmath,
amssymb,
aps,
prd,
]{revtex4-1}

\usepackage{times}
\usepackage{mathtools}
\usepackage{graphicx}
\usepackage{dcolumn}
\newcolumntype{d}[1]{D{.}{.}{#1}}
\usepackage{bm}
\usepackage{verbatim} 
\usepackage{cprotect} 
\usepackage{url}
\usepackage{subfigure} 
\usepackage{tikz}
\usepackage{afterpage}
\usepackage{nicefrac}
\usepackage{tabularx}
\newcolumntype{L}{>{\raggedright\arraybackslash}X}
\usepackage{multirow}
\usepackage[english]{babel}
\usepackage{natbib} 
\usepackage{glossaries} 
\usepackage{hyperref}
\usepackage{aas_macros} 
\usepackage{comment}

\graphicspath{{figures/}}

\usepackage{xcolor}

\begin{document}

\title{Investigating the effect of sensitivity of KAGRA on sky localization of gravitational-wave sources from compact binary coalescences}
\author{Alvin K. Y. Li}
\email{alvinli@g.ecc.u-tokyo.ac.jp}
\affiliation{RESCEU, The University of Tokyo, Tokyo, 113-0033, Japan}

\author{Peony K. K. Lai}
\email{peony.lai@link.cuhk.edu.hk}
\affiliation{Department of Physics, The Chinese University of Hong Kong}

\author{Elwin K. Y. Li}
\email{elwinlikayau@link.cuhk.edu.hk}
\affiliation{Department of Physics, The Chinese University of Hong Kong}

\author{Otto A. Hannuksela}
\affiliation{Department of Physics, The Chinese University of Hong Kong}
\date{\today}

\begin{abstract}
	The addition of KAGRA to the global gravitational-wave detector network introduces new baselines and complementary antenna response patterns. These features can enhance sky localization for compact binary coalescences. This study investigates KAGRA's role within the LIGO-Virgo-KAGRA network through a systematic injection analysis of binary neutron star signals. We construct sky localization maps using a radiometric, coherence-based framework. This approach enables isolation of the geometric and timing contributions of individual detectors. Localization performance is quantified by the fraction of events localized within $100~\mathrm{deg}^2$, the cumulative distribution of localization areas, and the median $90\%$ credible region. We also examine KAGRA's impact on the detection rate. By varying KAGRA's sensitivity across a wide range, we characterize its contribution in different regimes. As its sensitivity increases, KAGRA continues to improve sky localization. Even at its current sensitivity of approximately $10~\mathrm{Mpc}$, KAGRA provides measurable gains by breaking degeneracies through additional baselines and directional constraints. As sensitivity improves, KAGRA increases the signal-to-noise ratio and timing precision. This results in substantial reductions in the localization area. We identify a binary neutron star range of approximately $30~\mathrm{Mpc}$ as a practical benchmark for systematically suitable localization performance for electromagnetic follow-up. This value should be regarded as a conservative estimate rather than a strict threshold. In addition to improving localization, KAGRA increases the number of detectable events by enabling the detection of lower-signal-to-noise-ratio signals. Collectively, these effects enhance both the quantity and quality of gravitational-wave observations. The results demonstrate that even a relatively low-sensitivity detector can significantly improve network performance through geometric complementarity. This points to the importance of a geographically distributed detector network for multimessenger gravitational-wave astronomy.
\end{abstract}

\maketitle

\section{Introduction}\label{Section: Introduction}

The LIGO-Virgo-KAGRA (LVK) Collaboration has detected $\mathcal{O}(100)$ gravitational-wave (GW) events since the first direct observation in 2015~\cite{LIGOScientific:2016aoc,LIGOScientific:2025slb,LIGOScientific:2025hdt,LIGOScientific:2025yae,LIGOScientific:2025pvj,LIGOScientific:2025jau}.
The current global network consists of four interferometers: LIGO Hanford (H1), LIGO Livingston (L1), Virgo (V1), and KAGRA (K1)~\cite{LIGOScientific:2014pky,Virgo:2026fpe,KAGRA:2020tym,Aso:2013eba}.

Their geographical separation and relative orientations enable coherent measurements of GW signals.
These measurements form the basis for sky localization using triangulation and signal consistency~\cite{Fairhurst:2009tc,Fairhurst:2010is,2013CQGra..30o5004R}.
\begin{figure}[h!]
\centering
\includegraphics[width=\columnwidth]{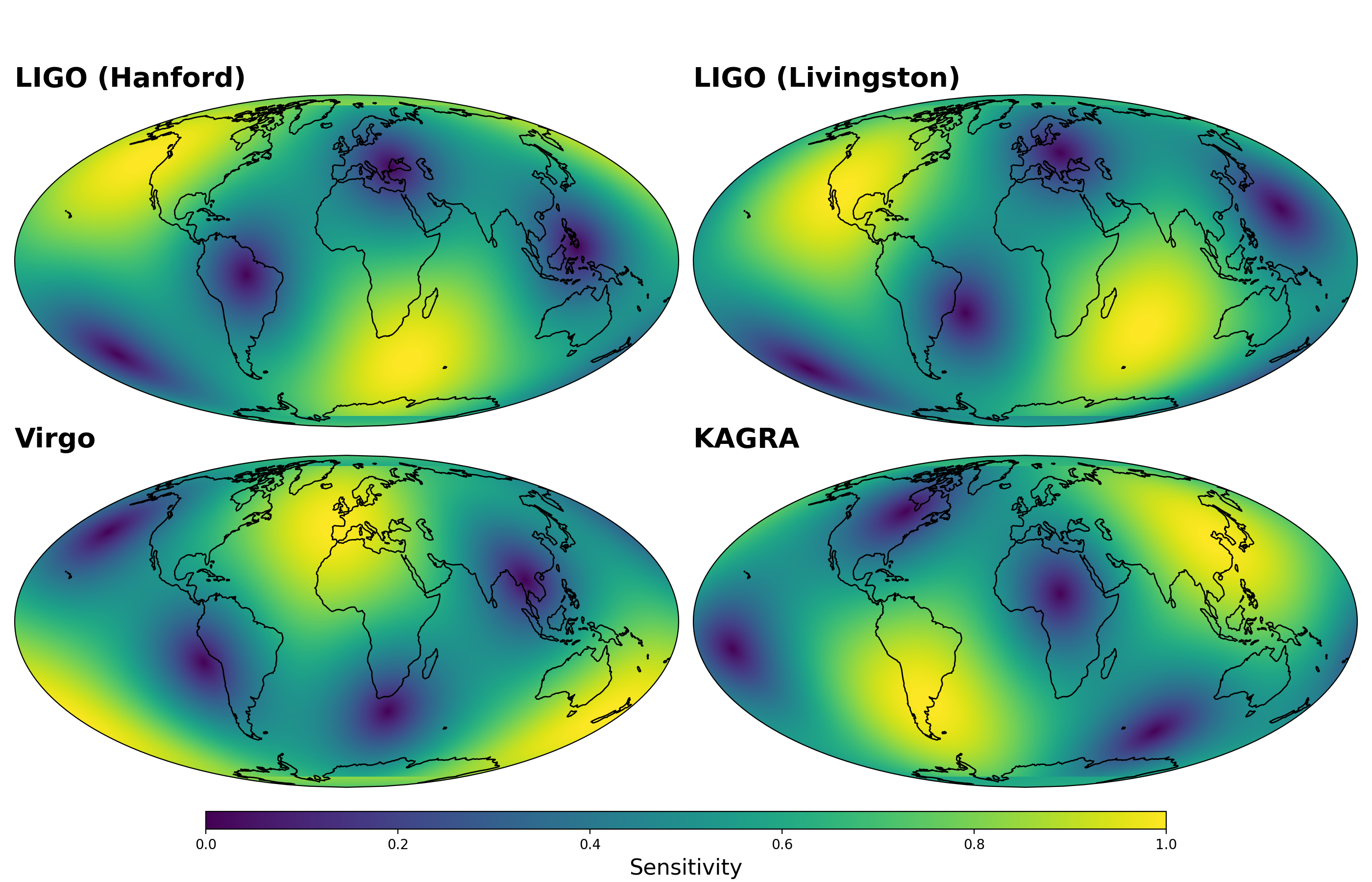}
\caption{Antenna response patterns of the LVK detectors across the sky, shown in equatorial coordinates. The color scale indicates relative sensitivity to GW signals. Each detector exhibits a characteristic quadrupolar response with regions of high sensitivity and nulls. KAGRA's antenna pattern differs significantly from those of the LIGO and Virgo detectors due to its geographic location and orientation. In particular, it provides enhanced sensitivity in regions of the sky where other detectors have reduced response (highlighted by dashed circles). Thus, it contributes complementary directional information to the network.
}
\label{fig:detector_network}
\end{figure}
The directional sensitivity of each detector is characterized by its antenna response pattern, which varies across the sky due to orientation and location~\cite{Schutz:2011tw}. As shown in Fig.~\ref{fig:detector_network}, while the LIGO and Virgo detectors exhibit broadly similar structures. However, KAGRA provides sensitivity in regions of the sky where the other detectors have reduced response. This complementarity is crucial for sky localization. Even if a detector has lower overall sensitivity, its ability to constrain or exclude specific sky regions can significantly improve localization. This improvement occurs through additional baselines and directional information.

Unlike electromagnetic (EM) observations, GW detections do not yield precise source positions. Instead, they produce probability sky maps whose size and morphology depend on the signal-to-noise ratio (SNR), detector geometry, and waveform bandwidth. Together, these factors determine the precision of timing, phase, and amplitude measurements across the network, and thus directly control localization accuracy~\cite{Fairhurst:2009tc,Singer:2015ema,Fairhurst:2010is,Wen:2010cr}.

Multimessenger events such as binary neutron star (BNS) mergers are of particular importance because they enable joint GW-EM observations that probe astrophysics, nucleosynthesis, and cosmology~\cite{Nissanke:2013fka,Metzger:2019zeh}. However, effective EM follow-up requires sufficiently small localization regions, typically $\lesssim 100~\mathrm{deg}^2$~\cite{Magee:2022kkc,Sachdev:2020lfd,KAGRA:2013rdx,Nissanke:2012dj,Chaudhary:2023vec,Ouzriat:2025ben,Cannon:2011vi}.

In addition to Bayesian methods such as Bayestar~\cite{Singer:2015ema,Singer:2014qca}, radiometry and coherence-based approaches~\cite{Tsutsui:2020sml} have been developed. These exploit detector cross-correlations to generate sky maps directly from data and provide complementary insight into how detector baselines and signal bandwidth affect angular resolution.

KAGRA introduces new baselines and orientations to the detector network, thereby improving localization~\cite{Tsutsui:2020sml}. Its geographic position and antenna pattern enhance sensitivity to regions of the sky less accessible to other detectors. Additional KAGRA baselines sharpen inter-detector time-delay and phase difference measurements, helping to resolve degeneracies in smaller networks~\cite{Grover:2013sha}.

A prominent example is GW170817, localized to $28~\mathrm{deg}^2$~\cite{LIGOScientific:2017vwq,LIGOScientific:2017ync}. While a network SNR of 32.4 aided precise timing, the key localization gain came from adding Virgo. Although Virgo was less sensitive, its non-detection indicated the source was in a sky region with weak antenna response, effectively ruling out large sky areas and breaking degeneracies in the two-detector (HL) setup, thus reducing the localization area~\cite{LIGOScientific:2017vwq,LIGOScientific:2017ync}. This case highlights how detector placement and response shape network performance.

Similar studies~\cite{Emma:2024mjs} using full parameter estimation offers a detailed statistical treatment of the signal and detector response. Studies of detector network performance have also shown that the inclusion of a full interferometer network improves localization through increased baseline coverage and duty cycle~\cite{Pankow:2019oxl}. For multimessenger astronomy, rapid sky localization is critical to enable timely electromagnetic follow-up. Our approach therefore attempts to align with low-latency methods, employing a radiometric, coherence-based framework~\cite{Tsutsui:2020sml} that provides a complementary and physically transparent way to investigate the roles of detector geometry and sensitivity.

The remainder of this paper is organized as follows. 
In Section~\ref{sec:method}, we describe the radiometric, coherence-based framework and the injection setup. 
In Section~\ref{sec:results}, we present the results on localization performance and its dependence on KAGRA sensitivity. 
Finally, in Section~\ref{sec:conclusion}, we summarize our findings and discuss their implications for multimessenger gravitational-wave astronomy.

\section{Methodology}\label{sec:method}
\subsection{Approximate Radiometric approach for localization}\label{subsec:radiometric}
In this work, we generate GW skymaps using a radiometric, coherence-based framework that measures signal consistency across detector networks as a function of sky position~\cite{Klimenko:2008fu,Tsutsui:2020sml}.
Bayestar~\cite{Singer:2015ema} provides rapid and accurate localization by incorporating timing, phase, and amplitude information within a calibrated likelihood model. However, marginalization over distance, inclination, and polarization remains computationally expensive, particularly for studies focused on isolating the individual roles of detector baselines, antenna patterns, and sensitivity. Rather than replacing Bayestar, this study employs a complementary framework that is relatively fast, physically interpretable, and well-suited to systematic sensitivity analyses. This approach enables direct examination of how weakly responding or non-detecting detectors constrain sky position through network geometry, which is central to understanding the role of KAGRA.

We model the GW signal in the frequency domain. For each detector $I$, the detector response to a given signal $h$ is given by~\cite{Allen:2005fk,Jaranowski:1998qm}
\begin{align}
	s_I(f, \Omega, \Psi) = F_{I,+} (\Omega, \Psi) h_+(f) + F_{I,\times} (\Omega, \Psi) h_\times(f),
\end{align}
where $F_{I,+}$ and $F_{I,\times}$ are the antenna response functions evaluated at sky location $\Omega$ with polarization angle $\Psi$. $h_+(f)$ and $h_\times(f)$ are the plus and cross polarization waveforms, which are generated using standard waveform approximants TaylorF2~\cite{Buonanno:2009zt} and scaled with luminosity distance to the source as 
\begin{align}
	h(f) \propto \frac{1}{d}.
\end{align}
We assume each detector measures a GW signal plus stationary Gaussian noise. In frequency space, detector $I$'s data is $d_I(f) = h_I(f) + n_I(f)$, with $n_I(f)$ zero-mean Gaussian noise of one-sided PSD $S_I(f)$. Noise between detectors is uncorrelated: $\langle n_I(f), n_J(f') \rangle = \frac{1}{2} S_I(f) \delta_{IJ}\delta(f-f')$.
Under these assumptions, the Gaussian likelihood for a signal hypothesis~\cite{Cutler:1994ys,Veitch:2014wba,Ashton:2018jfp,Poisson:1995ef} is
\begin{align}
	p(d|h) \propto \exp\left[-\frac{1}{2}\sum_I \langle d_I-h_I \mid d_I-h_I \rangle_I \right],
\end{align}
where the detector-weighted inner product is
\begin{align}
	\langle a \mid b \rangle_I = 4\,\Re \int_0^\infty \frac{a_I^*(f)\,b_I(f)}{S_I(f)}\,df .
\end{align}
Equivalently, the log-likelihood is
\begin{align}
	\ln p(d|h) = \mathrm{const} + \sum_I \langle d_I \mid h_I \rangle_I - \frac{1}{2}\sum_I \langle h_I \mid h_I \rangle_I.
\end{align}

For localization, the signal model depends on sky direction through the detector antenna responses and through the arrival time at each detector. For a fixed arrival time, the detector strain in detector $I$ is related to the underlying response $s_I$ by a phase factor,
\begin{align}
	h_I(f, \Omega) = s_I(f, \Omega)\,\exp\!\left[-2\pi i f \tau_I(\Omega)\right],
\end{align}
where $\tau_I(\Omega)$ is the geometric arrival time in detector $I$ for sky direction $\Omega$. The relative arrival time between two detectors $I$ and $J$ is~\cite{Tsutsui:2020sml,Fairhurst:2009tc,Fairhurst:2010is}
\begin{align}
	\tau_{IJ}(\Omega) = \tau_J(\Omega)- \tau_I(\Omega) = \frac{(\mathbf{r}_J-\mathbf{r}_I)\cdot \mathbf{n}(\Omega)}{c},
\end{align}
where $\mathbf{r}_I$ is the position vector of detector $I$, $\mathbf{n}(\Omega)$ is the unit vector pointing toward sky location $\Omega$, and $c$ is the speed of light.

Denoting the true source direction by $\Omega_{\mathrm{true}}$, we define the delay residual for a trial sky position $\Omega$ as
\begin{align}
	\Delta \tau_{IJ}(\Omega) = \tau_{IJ}(\Omega)-\tau_{IJ}(\Omega_{\mathrm{true}}).
\end{align}
At this point, a full Bayesian treatment would evaluate the single-detector matched-filter terms and marginalize over the unknown extrinsic parameters, including distance, inclination, polarization and reference phase~\cite{Singer:2015ema,Veitch:2014wba,Ashton:2018jfp}. In this work, rather than imposing that full marginalization, we construct an approximate statistic that isolates the leading sky-dependent effect arising from relative timing consistency across the detector network.

Consider the product of the model strains in two detectors,
\begin{align}
	h_I(f) h_J^*(f)
	&=
	\left[F_{I,+}h_+ + F_{I,\times}h_\times\right]
	\left[F_{J,+}h_+ + F_{J,\times}h_\times\right]^* \nonumber \\
	&\quad \times \exp\left[-2\pi i f (\tau_I - \tau_J)\right].
\end{align}
The sky dependence {\bf associated with relative arrival time} is therefore contained in the complex phase factor~\cite{Singer:2015ema}
\begin{align}
	\exp\left[-2\pi i f (\tau_I - \tau_J)\right] = \exp\left[2\pi i f \tau_{IJ}\right]
\end{align}
with $\tau_{IJ} = \tau_J - \tau_I$. If we compare a trial sky position $\Omega$ to the true direction $\Omega_{\mathrm{true}}$, the residual phase shift is then $\exp\left[2\pi i f \Delta\tau_{IJ}\right]$.
Thus, when the trial sky position is correct, the phase is zero, and the two detector responses are maximally coherent; when the sky position is incorrect, the residual phase oscillates with frequency and suppresses the coherent sum.

Motivated by this structure, we define a detector-pair coherence integral by weighting the phase factor over frequency~\cite{Ballmer:2005uw,Tsutsui:2020sml,PhysRevD.75.123003,Thrane:2013oya}
\begin{align}
	\widetilde{C}_{IJ}(\Omega)
	=
	\int_0^\infty df\, \widehat{w}_{IJ}(f)\,
	\exp\left[2\pi i f \Delta\tau_{IJ}\right] ,
\end{align}
where $\widehat{w}_{IJ}(f)$ is a normalized non-negative frequency weight~\cite{Cutler:1994ys,Allen:2005fk} satisfying
\begin{align}
	\int_0^\infty df\, \widehat{w}_{IJ}(f)=1 .
\end{align}
In our implementation, we take
\begin{align}
	\widehat{w}_{IJ}(f)
	\propto
	\frac{|h_I(f)|\,|h_J(f)|}{S_I(f)\,S_J(f)} ,
\end{align}
followed by normalization. This choice emphasizes frequencies at which both detectors have appreciable signal power relative to the noise, while keeping the statistic simple and computationally efficient.

The quantity $\widetilde{C}_{IJ}(\Omega)$ is in general complex. Writing it in terms of real and imaginary parts,
\begin{align}
	\widetilde{C}_{IJ}(\Omega)
	&=
	\int df\,\widehat{w}_{IJ}(f)\cos\!\left(2\pi f\,\Delta\tau_{IJ}(\Omega)\right)\\
	&+
	i \int df\,\widehat{w}_{IJ}(f)\sin\!\left(2\pi f\,\Delta\tau_{IJ}(\Omega)\right) .
\end{align}
Rather than retaining the full complex coherence, we keep only its real part and define
\begin{align}
	C_{IJ}(\Omega)
	\equiv
	\Re \widetilde{C}_{IJ}(\Omega)
	=
	\int df\,\widehat{w}_{IJ}(f)\cos\!\left(2\pi f\,\Delta\tau_{IJ}(\Omega)\right) .
\end{align}
The cosine term isolates the component of the coherence that is even under $\Delta\tau_{IJ}\rightarrow -\Delta\tau_{IJ}$ and depends only on timing consistency~\cite{Tsutsui:2020sml,Ballmer:2005uw}. By contrast, the sine term retains information about the signed phase offset and is therefore more sensitive to additional phase structure that, in a full treatment, would couple to the unknown coalescence phase, polarization, and detector response model. In addition, the imaginary component does not constructively combine across frequencies in the absence of a consistent phase model, and is therefore more susceptible to cancellation under realistic uncertainties in phase, polarization, and calibration. Since our goal here is not to reconstruct the full complex matched-filter likelihood but to build a quick approximation driven primarily by relative arrival times, we neglect the imaginary part and use only the cosine projection~\cite{Nishizawa:2009bf,Tsutsui:2020sml}. Equivalently, this can be viewed as assuming that the dominant sky-dependent information is contained in the time-alignment penalty rather than in the signed complex phase itself.

This approximation is also consistent with the behavior near the true sky location~\cite{Fairhurst:2009tc,Fairhurst:2010is}. When $\Delta\tau_{IJ}$ is small,
\begin{align}
	\exp\left[2\pi i f\,\Delta\tau_{IJ}\right]
	&\approx
	1 + 2\pi i f\,\Delta\tau_{IJ}
	-\frac{1}{2}(2\pi f\,\Delta\tau_{IJ})^2\\
	& \quad + \mathcal{O}(\Delta\tau_{IJ}^3) + ...
\end{align}
so that
\begin{align}
	\cos(2\pi f\,\Delta\tau_{IJ})
	=
	1-\frac{1}{2}(2\pi f\,\Delta\tau_{IJ})^2 + \mathcal{O}(\Delta\tau_{IJ}^4).
\end{align}
The leading deviation from perfect coherence is therefore quadratic in the timing error, which makes the cosine a natural measure of mismatch. By contrast, the sine term is odd in $\Delta\tau_{IJ}$ and vanishes at the true sky location to first order only after averaging over symmetric phase uncertainties or when the signed phase information is not modeled explicitly. For a timing-dominated localization statistic, keeping the cosine term therefore captures the leading penalty associated with an incorrect sky position.

Using this pairwise coherence, we define a global mismatch statistic over the detector network,
\begin{align}
	\Delta \Lambda(\Omega)
	=
	\sum_{I<J} \lambda_{IJ}\,\left[1-C_{IJ}(\Omega)\right],
\end{align}
where $\lambda_{IJ}$ is a non-negative weight assigned to baseline $(I,J)$. In practice, these weights may be chosen in several ways. They may be set uniformly, scaled by the weaker detector's SNR in the pair, scaled by the product of the detector SNRs, or chosen using timing-based weights proportional to the inverse combined timing variance. This freedom allows the statistic to emphasize either purely geometric baseline information or the expected statistical precision of the corresponding detector pair.

The quantity $1-C_{IJ}(\Omega)$ vanishes for a perfectly consistent sky direction and increases as the phase mismatch grows. The network statistic $\Delta\Lambda(\Omega)$ therefore plays the role of an approximate negative log-likelihood, constructed from the dominant timing-coherence contribution of each baseline. We then convert this mismatch into a skymap probability~\cite{Singer:2015ema,Tsutsui:2020sml} through
\begin{align}
	P(\Omega) \propto \exp\left[-\Delta\Lambda(\Omega)\right],
\end{align}
followed by normalization over all sky pixels~\cite{Gorski:2004by},
\begin{align}
	\sum_{\Omega} P(\Omega) = 1 .
\end{align}
Credible regions are computed by ranking pixels in descending probability and accumulating probability until the desired confidence level is reached.

This construction should be viewed as an approximate radiometric localization method rather than a full coherent Bayesian likelihood. The key approximations are: (i) reducing complex coherence to its timing-sensitive real component, (ii) neglecting explicit modeling of phase, polarization, and amplitude correlations~\cite{Nishizawa:2009bf,Tsutsui:2020sml}, and (iii) avoiding marginalization over extrinsic parameters~\cite{Tsutsui:2020sml}. These simplifications significantly reduce computational cost while preserving a direct and physically transparent connection between sky localization and detector geometry, bandwidth, and sensitivity. In contrast to Bayestar, which evaluates a calibrated likelihood and marginalizes over extrinsic parameters, our statistic isolates the leading timing-coherence contribution and treats other effects in an averaged or implicit manner.

\subsection{Injections and Setup}

To quantify KAGRA's contribution to sky localization in the LVK network, we perform an injection study using simulated BNS signals. Sky localization maps are generated using the radiometric method described in Sec.~\ref{subsec:radiometric}.

\subsubsection{Injection population}

We generate a total of 10000 compact binary coalescence signals with component masses fixed at $1.4\,M_{\odot}$--$1.4\,M_{\odot}$, representative of typical BNS systems. The sources are distributed uniformly in comoving volume~\cite{KAGRA:2021duu,LIGOScientific:2020kqk,LIGOScientific:2017vwq}, such that the luminosity distance follows $p(d) \propto d^2$, within the range $d \in [40,\,200]\,\mathrm{Mpc}$. This range is chosen to probe both well-localized and marginal detections relative to the detector sensitivities.

The extrinsic parameters of each injection are randomly sampled. Sky locations are distributed isotropically over the celestial sphere, inclination angles are drawn uniformly in $\cos\iota \in [-1,1]$, and polarization angles $\psi$ and coalescence phases $\phi_c$ are sampled uniformly in $[0,\,2\pi)$. The coalescence time $t_c$ is sampled uniformly over one sidereal day,
\begin{align}
	t_c \sim \mathrm{Uniform}(t_{\mathrm{ref}},\, t_{\mathrm{ref}} + T_{\mathrm{sidereal}}),
\end{align}
where $T_{\mathrm{sidereal}} = 86164~\mathrm{s}$. This procedure effectively randomizes the detector orientation relative to the sky through Earth's rotation, ensuring that antenna pattern effects are properly averaged and that no preferred sky orientation is introduced.

The component spins are assumed to be aligned with the orbital angular momentum, with dimensionless spin magnitudes sampled uniformly in the range $\chi_{1z}, \chi_{2z} \in [-0.1,\,0.1]$, while in-plane spin components are set to zero. All signals are generated using the TaylorF2 waveform approximant~\cite{Buonanno:2009zt,Sathyaprakash:2009xs}.

\subsubsection{Detector configuration and sensitivity modeling}

We consider a reference detector network consisting of H1, L1, and V1, and a test network that includes K1 in addition. The detector sensitivities are modeled using PSD representative of the O4a configuration for LIGO~\cite{LIGOScientific:2025gwtc4_sensitivity,LIGOScientific:2025snk}, the O3 configuration for Virgo~\cite{LIGO:T2200043,Virgo:2022ysc}, and O4 simulation estimate for KAGRA at $10$Mpc~\cite{LIGO:T2200043,KAGRA:2025dra}. 

To quantify the impact of KAGRA sensitivity, we scale the amplitude spectral density (ASD) of KAGRA by a factor $g$, such that
\begin{align}
	S_{K1}(f) \rightarrow g^2 S_{K1}(f),
\end{align}
where $S(f)$ denotes the PSD. We explore a wide range of ASD scaling factors spanning more than an order of magnitude, corresponding to different hypothetical KAGRA sensitivities.

\begin{figure}[h!]
\centering
\includegraphics[width=\columnwidth]{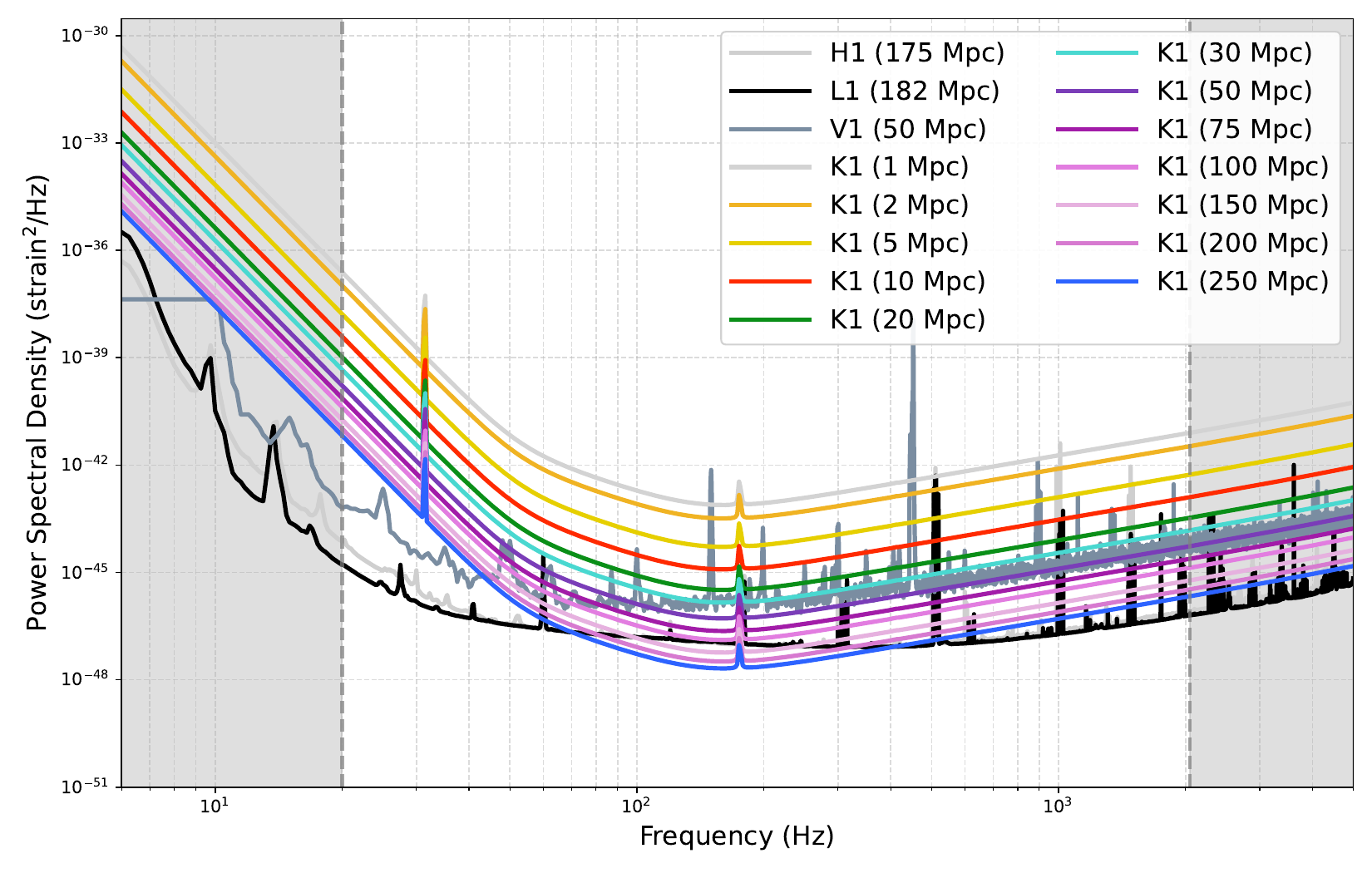}
\caption{
PSDs of the detector network used in this study. H1, L1 and V1 PSDs are fixed, while the KAGRA (K1) PSD is scaled to represent different sensitivity levels. Each KAGRA curve corresponds to a different effective BNS range, spanning from $1$~Mpc to $250$~Mpc. Scaling the ASD by a factor $g$ corresponds to scaling the PSD by $g^2$, enabling a systematic study of the impact of KAGRA sensitivity on sky localization.
}
\label{fig:psd_scaling}
\end{figure}

Figure~\ref{fig:psd_scaling} shows the resulting PSDs for the detector network. While the PSDs of H1, L1, and V1 are held fixed, the KAGRA PSD is systematically varied, corresponding to effective BNS ranges from $1$~Mpc to $250$~Mpc. For reference, the current KAGRA sensitivity is approximately $10$~Mpc, placing it toward the lower end of the explored range. This provides a continuous interpolation between a nearly non-contributing detector and one with sensitivity comparable to that of the LIGO and Virgo detectors.

\subsubsection{Analysis}
For each realization and ASD scaling, matched filtering is performed to compute the SNR in each detector. Events are required to satisfy a network SNR threshold of $\rho_{\mathrm{net}} \geq 8$ in the reference network.

Each signal is embedded in stationary Gaussian noise consistent with the detector PSDs. Matched filtering is then used to identify candidate triggers and extract the corresponding timing and SNR information in each detector.

This setup allows us to isolate the role of detector sensitivity in sky localization. At low sensitivity, KAGRA primarily contributes through geometric baseline information with large timing uncertainty, while at higher sensitivity, it provides stronger timing constraints and additional coherence, leading to improved localization.

To account for finite measurement precision, we include detector timing noise in the analysis. For each detector $I$, a timing perturbation is drawn from a Gaussian distribution,
\begin{align}
	dt_I \sim \mathcal{N}(0, \sigma_{t,I}^2),
\end{align}
where the timing uncertainty~\cite{Fairhurst:2009tc,Fairhurst:2010is,Finn:1992wt,Vallisneri:2007ev} is approximated as
\begin{align}
	\sigma_{t,I} \approx \frac{1}{2\pi \rho_I \sigma_{f,I}},
\end{align}
with $\rho_I$ the detector SNR and $\sigma_{f,I}$ the effective signal bandwidth. These perturbations are incorporated into the inter-detector delays via
\begin{align}
	\Delta\tau_{IJ} \rightarrow \Delta\tau_{IJ} - (dt_J - dt_I),
\end{align}
thereby modeling the impact of timing uncertainty on sky localization.

For each injection, sky localization probability maps are constructed using the radiometric coherence-based method~\cite{Tsutsui:2020sml}. This method evaluates the consistency of inter-detector timing across the sky by computing pairwise coherence between detectors and forming a network mismatch statistic. The mismatch is then mapped to a normalized probability distribution over sky pixels, producing the final skymap for each event.

This injection framework enables a systematic study of how detector sensitivity and network configuration affect sky localization performance, with particular emphasis on KAGRA's role in constraining source position via additional baselines and timing information.

The full set of injection parameters and their distributions are summarized in Table~\ref{tab:injection_parameters}.

\begin{table}[t]
\centering
\caption{Injection parameter distributions for the simulated BNS population.}
\begin{tabular}{ll}
\hline
\textbf{Parameter} & \textbf{Distribution / Value} \\
\hline
Component masses $(m_1, m_2)$ & $(1.4,\,1.4)\,M_{\odot}$ (fixed) \\
Waveform model & TaylorF2 \\
Luminosity distance $d$ & Uniform in volume, $p(d) \propto d^2$, \\
 & $d \in [40,\,200]\,\mathrm{Mpc}$ \\
Right ascension $\alpha$ & Uniform in $[0,\,2\pi)$ \\
Declination $\delta$ & Isotropic (uniform in $\sin\delta$) \\
Inclination $\iota$ & Uniform in $\cos\iota \in [-1,1]$ \\
Polarization $\psi$ & Uniform in $[0,\,2\pi)$ \\
Coalescence phase $\phi_c$ & Uniform in $[0,\,2\pi)$ \\
Spin components $(\chi_{1z}, \chi_{2z})$ & Uniform in $[-0.1,\,0.1]$ \\
In-plane spins & Zero \\
Reference frequency $f_{\mathrm{ref}}$ & $0$ Hz \\
Coalescence time $t_c$ & Uniform over one sidereal day \\
\hline
\end{tabular}
\label{tab:injection_parameters}
\end{table}

\section{Results}\label{sec:results}

\subsection{Event classification: HLV-detectable vs HLVK-detectable}

To isolate the impact of KAGRA on sky localization, we subdivide the simulated events into two categories based on their detectability:
\begin{enumerate}
	\item[(1)] \textbf{HLV-detectable events}: Events that satisfy the network SNR threshold $\rho_{\mathrm{net}}^{\mathrm{HLV}} \geq 8$ using only H1, L1 and V1 detectors.
	\item[(2)] \textbf{HLVK-detectable events}: Events that satisfy $\rho_{\mathrm{net}}^{\mathrm{HLVK}} \geq 8$ when KAGRA (K1) is included, but may not meet the detection threshold in the HLV network alone.
\end{enumerate}

This classification allows us to disentangle two distinct roles of KAGRA: improving localization for already-detectable events, and increasing the number of detectable events in the network.

\subsection{Overview of localization performance}

We quantify localization performance using the fraction of events with $90\%$ credible area $A_{90} \leq 100~\mathrm{deg}^2$~\cite{Magee:2022kkc,Sachdev:2020lfd,KAGRA:2013rdx,Nissanke:2012dj}, a commonly adopted threshold for efficient EM follow-up.

For each event, $A_{90}$ is computed as the smallest sky region containing $90\%$ of the posterior probability. This metric provides a direct measure of whether an event is practically localizable for multimessenger observations.

\subsection{Localization performance with KAGRA}

\begin{figure}[h!]
\centering
\includegraphics[width=\columnwidth]{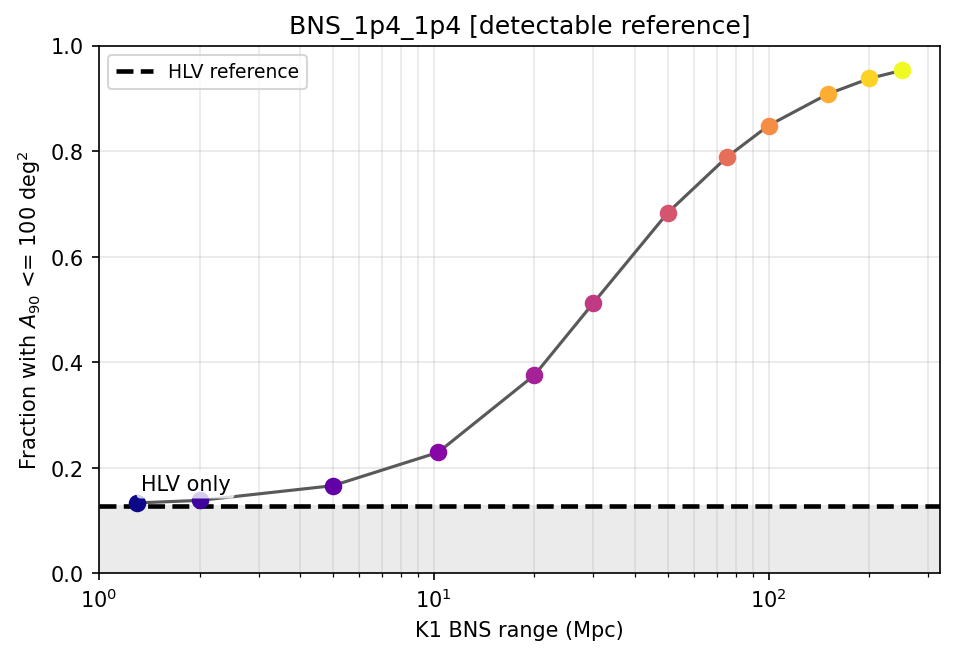}
\vspace{2mm}
\includegraphics[width=\columnwidth]{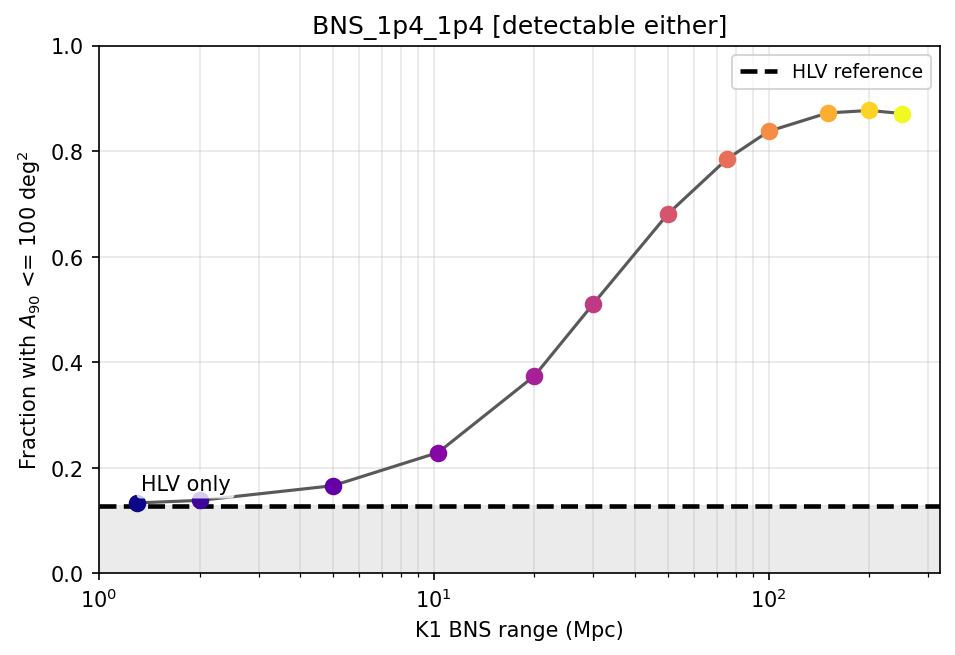}
\caption{
Fraction of events localized within $100~\mathrm{deg}^2$ as a function of KAGRA sensitivity.
Top: HLV-detectable events.
Bottom: HLVK-detectable events.
}
\label{fig:localization_fraction}
\end{figure}

Figure~\ref{fig:localization_fraction} shows the fraction of well-localized events as a function of KAGRA sensitivity.

For HLV-detectable events, the inclusion of KAGRA leads to a clear and systematic improvement. Even at the current sensitivity of $\sim 10~\mathrm{Mpc}$, the fraction of well-localized events increases from $\sim 13\%$ to $\sim 23\%$.

This improvement arises primarily from geometric effects. The addition of KAGRA introduces a new detector baseline, providing independent constraints on arrival time differences. Even when KAGRA contributes little SNR, this geometric information helps break degeneracies inherent in the HLV network~\cite{Fairhurst:2009tc,LIGOScientific:2017vwq,Fairhurst:2010is,Grover:2013sha}.

As KAGRA sensitivity increases, the improvement becomes more pronounced. The fraction of well-localized events rises rapidly, reaching $\sim 80\%$ at $\sim 80~\mathrm{Mpc}$ and exceeding $90\%$ at high sensitivities. This reflects the growing importance of precise timing and coherent signal reconstruction across the entire detector network.

For the HLVK-detectable population, a similar trend is observed. Although the inclusion of additional low-SNR events slightly broadens the distribution, the overall improvement in sensitivity remains strong and monotonic.

\subsection{Cumulative distribution of localization areas}

To understand the improvement across the full event population, we examine the cumulative distribution of localization areas.

\begin{figure}[h!]
\centering
\includegraphics[width=\columnwidth]{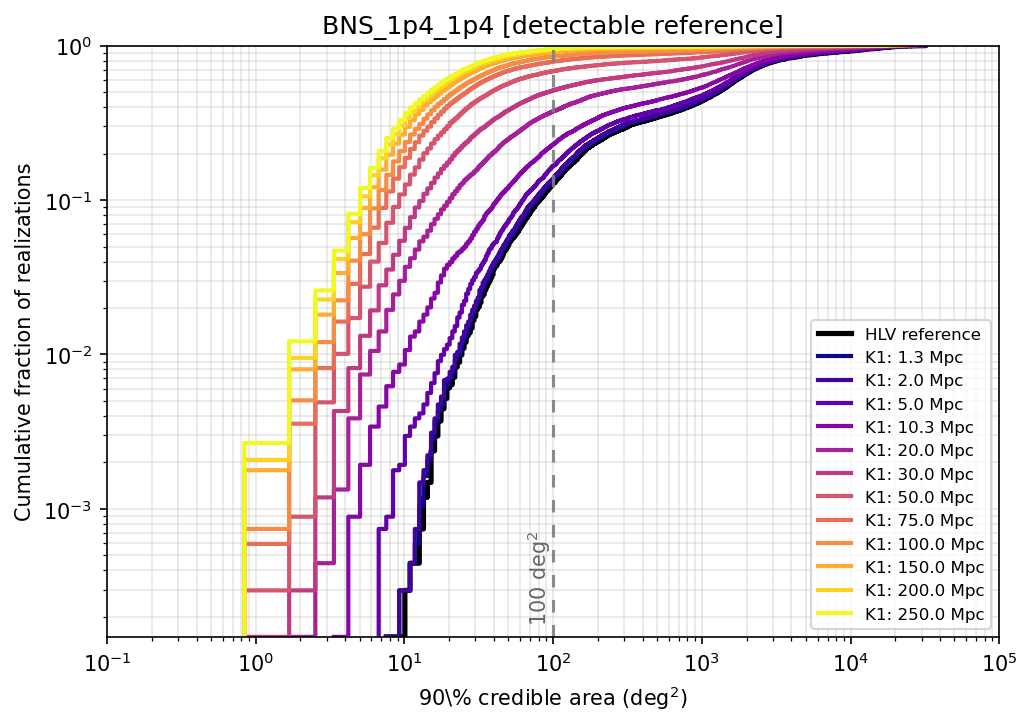}
\vspace{2mm}
\includegraphics[width=\columnwidth]{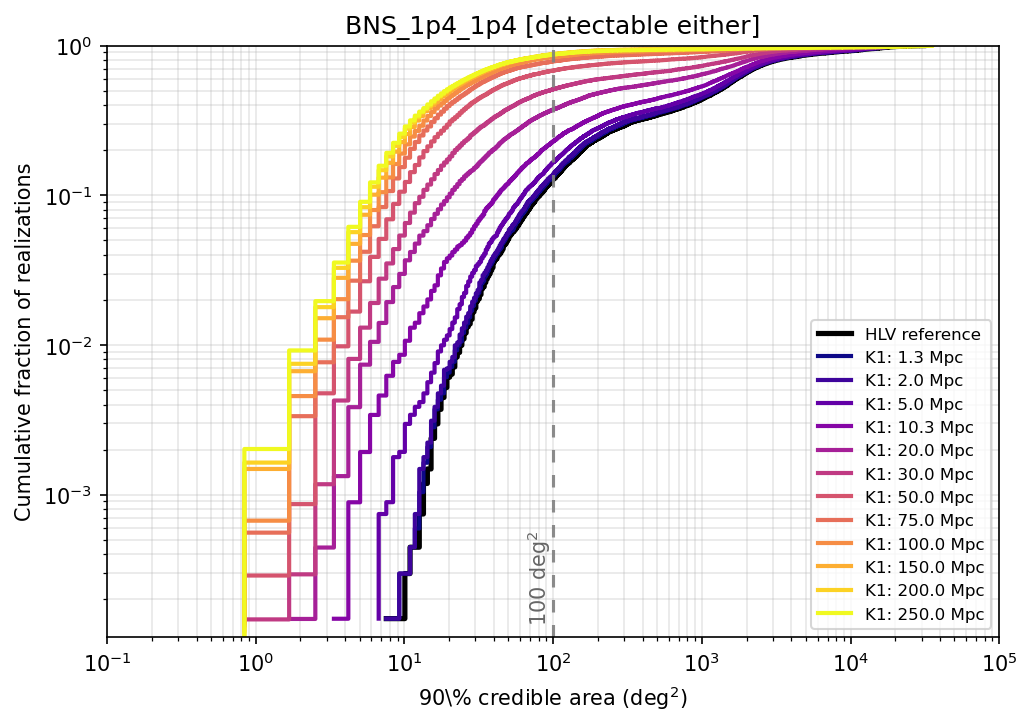}
\caption{
Cumulative distribution of $90\%$ credible areas. The vertical line marks $100~\mathrm{deg}^2$.
}
\label{fig:cumulative_distribution}
\end{figure}

As KAGRA sensitivity increases, the cumulative distribution shifts steadily toward smaller areas. This shift occurs across the entire range of $A_{90}$, indicating that the improvement is not limited to a subset of events but applies broadly.

In particular, the fraction of events below $100~\mathrm{deg}^2$ increases significantly, reinforcing the conclusions drawn from Fig.~\ref{fig:localization_fraction}.

\subsection{Impact on detection rate}

In addition to localization, KAGRA also affects the detection rate.

\begin{figure}[h!]
\centering
\includegraphics[width=\columnwidth]{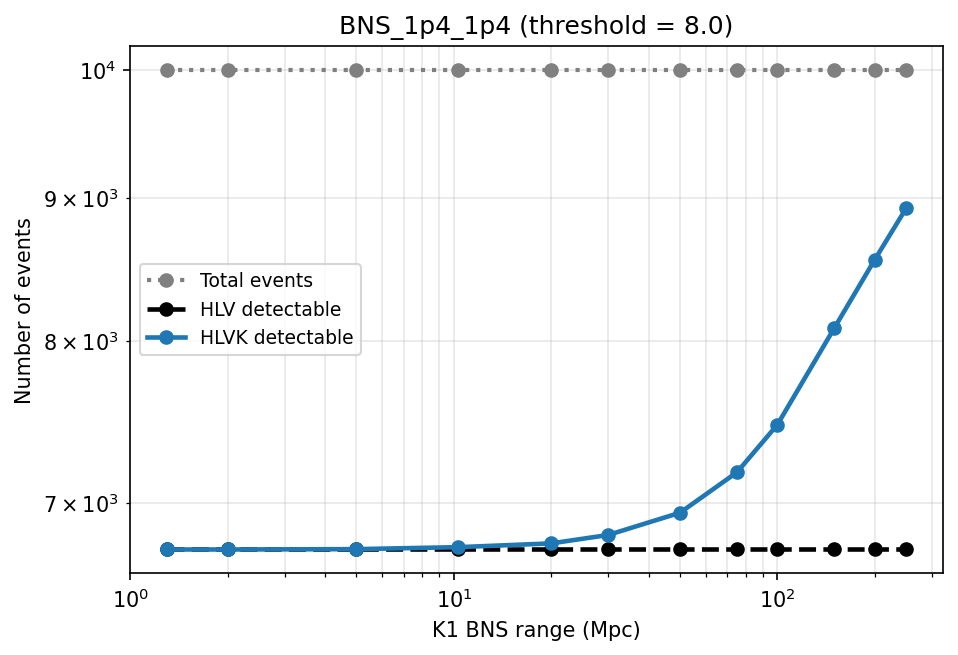}
\caption{
Number of detected events as a function of KAGRA sensitivity.
}
\label{fig:detection_counts}
\end{figure}

While the number of HLV-detectable events remains constant, the number of HLVK-detectable events increases significantly with KAGRA sensitivity. At high sensitivity, the detection rate increases by more than $\sim 30\%$.

This increase is primarily due to the inclusion of lower-SNR events that become detectable only when KAGRA contributes additional signal power.

\subsection{Median localization and target sensitivity}

\begin{figure}[h!]
\centering
\includegraphics[width=\columnwidth]{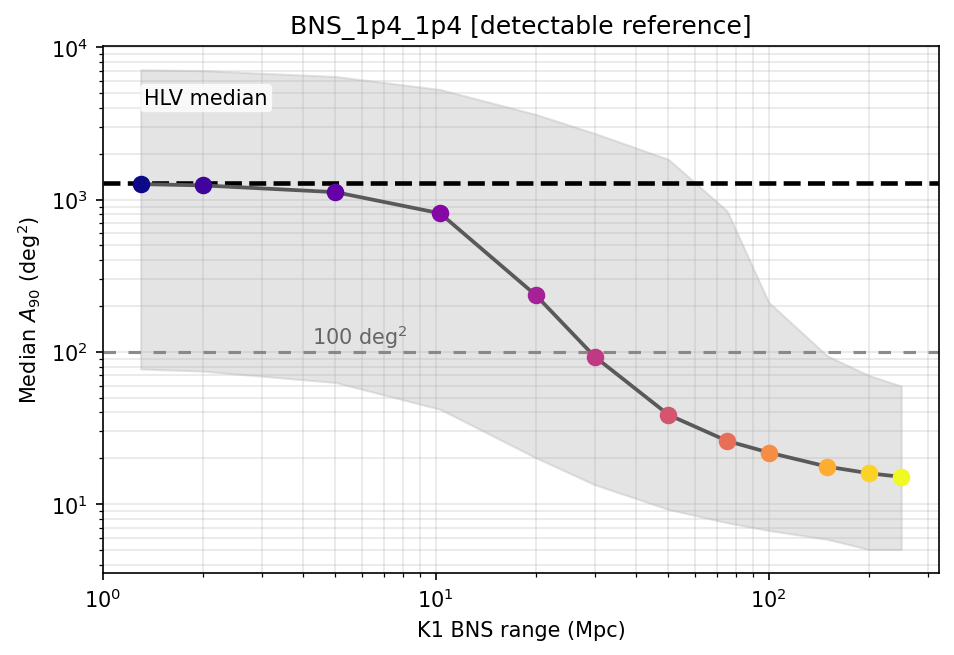}
\vspace{2mm}
\includegraphics[width=\columnwidth]{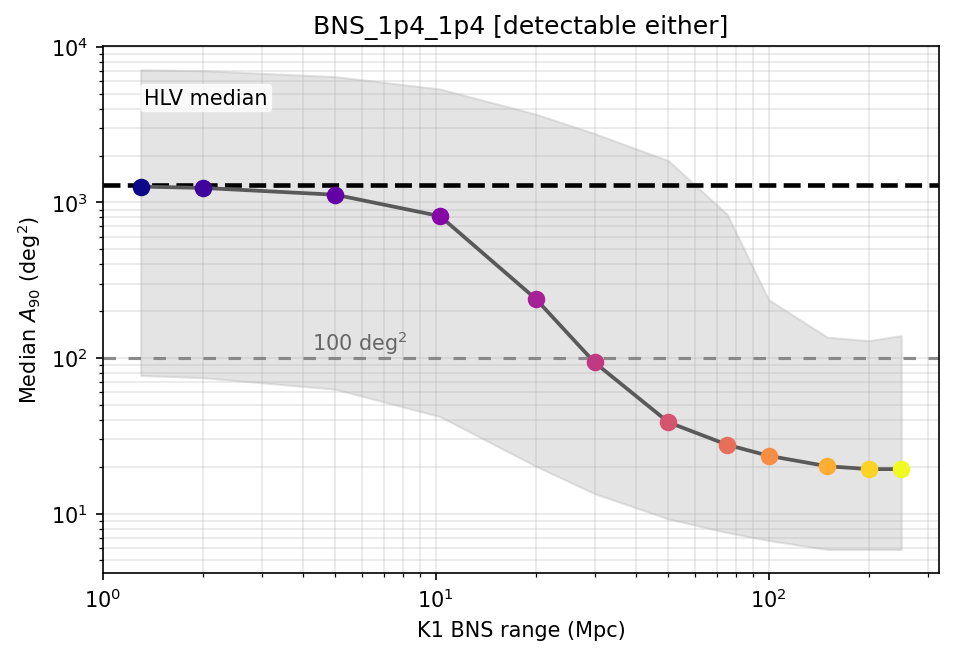}
\caption{
Median $90\%$ credible area as a function of KAGRA sensitivity.
}
\label{fig:target_sensitivity}
\end{figure}

Figure~\ref{fig:target_sensitivity} summarizes the typical localization performance through the median $A_{90}$.

The median localization area decreases dramatically with increasing sensitivity, from $\sim 10^3~\mathrm{deg}^2$ to $\sim 20~\mathrm{deg}^2$.

A key transition occurs around $\sim 30~\mathrm{Mpc}$, where the median crosses the $100~\mathrm{deg}^2$ threshold. This marks the regime in which localization becomes systematically useful for EM follow-up.

Importantly, this should be interpreted as a conservative benchmark. The improvement is continuous, and measurable gains are already present at the current KAGRA sensitivity of $\sim 10~\mathrm{Mpc}$.

\subsection{Evolution of localization with KAGRA sensitivity}

\begin{figure*}[t]
\includegraphics[width=0.32\textwidth]{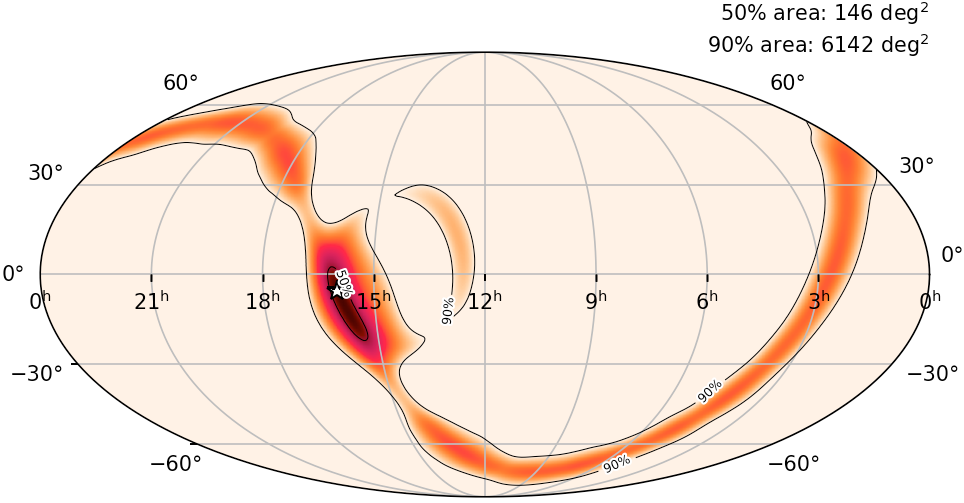}
\includegraphics[width=0.32\textwidth]{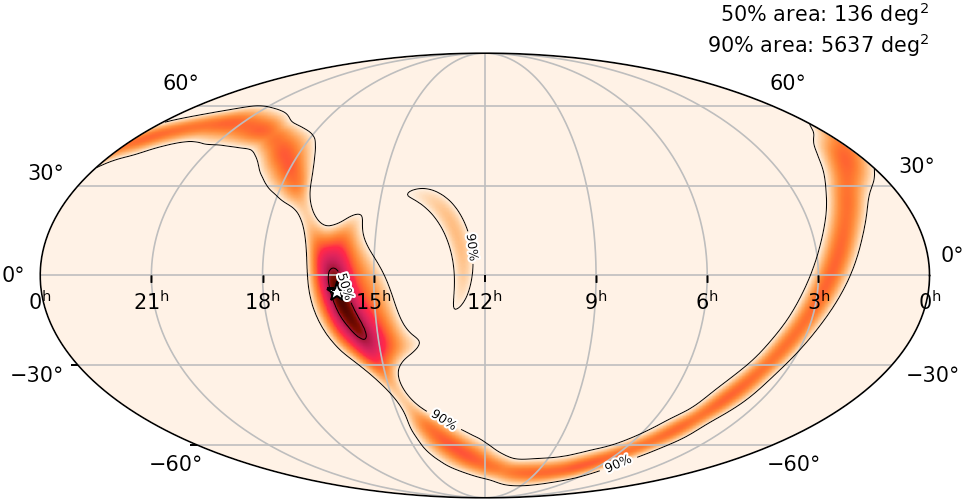}
\includegraphics[width=0.32\textwidth]{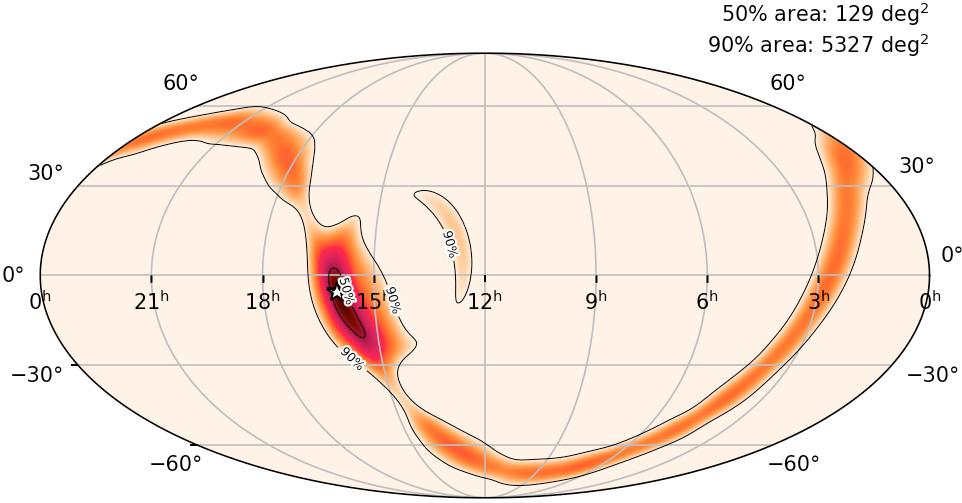}

\vspace{2mm}

\includegraphics[width=0.32\textwidth]{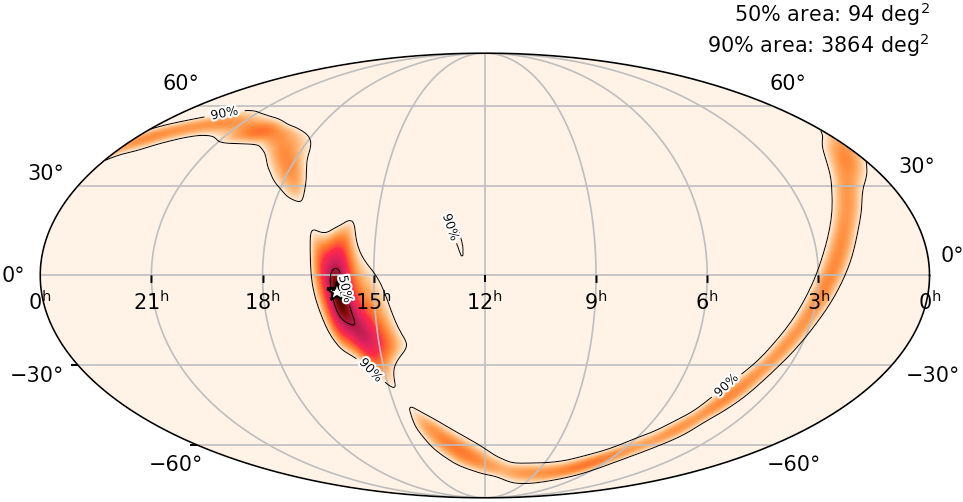}
\includegraphics[width=0.32\textwidth]{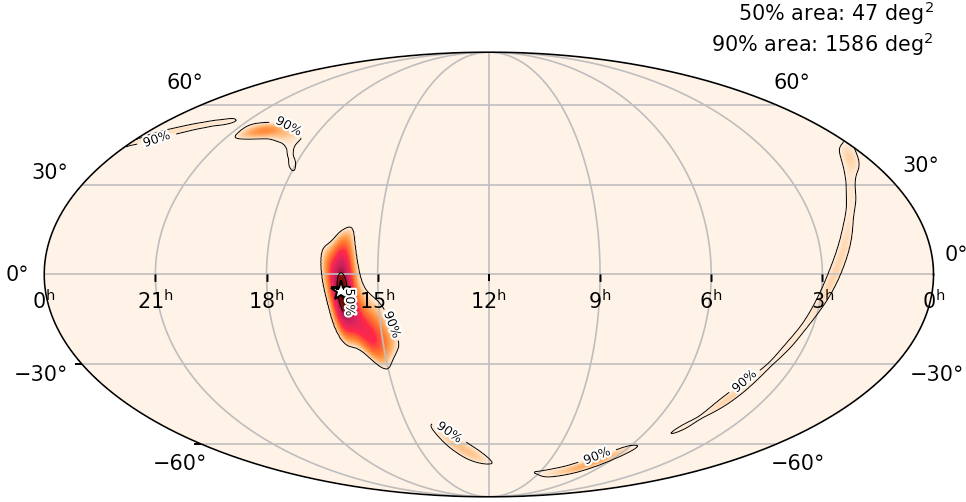}
\includegraphics[width=0.32\textwidth]{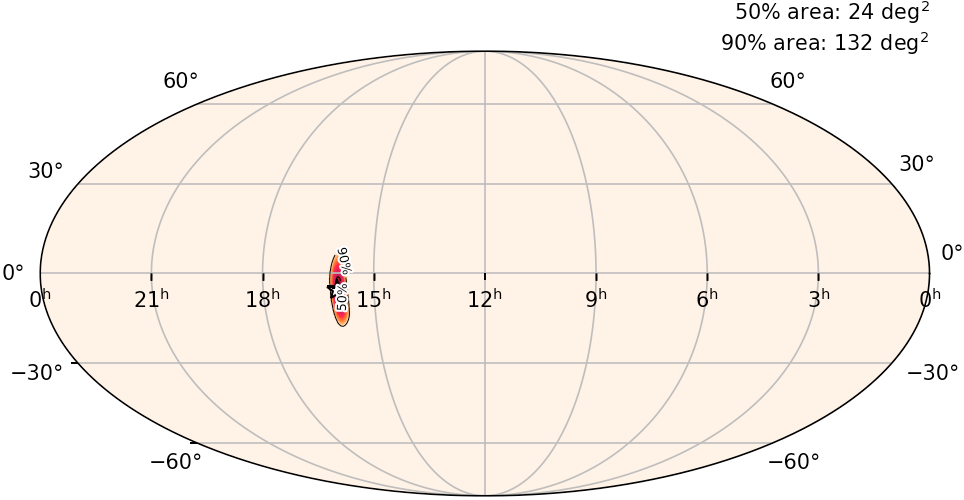}

\vspace{2mm}

\includegraphics[width=0.32\textwidth]{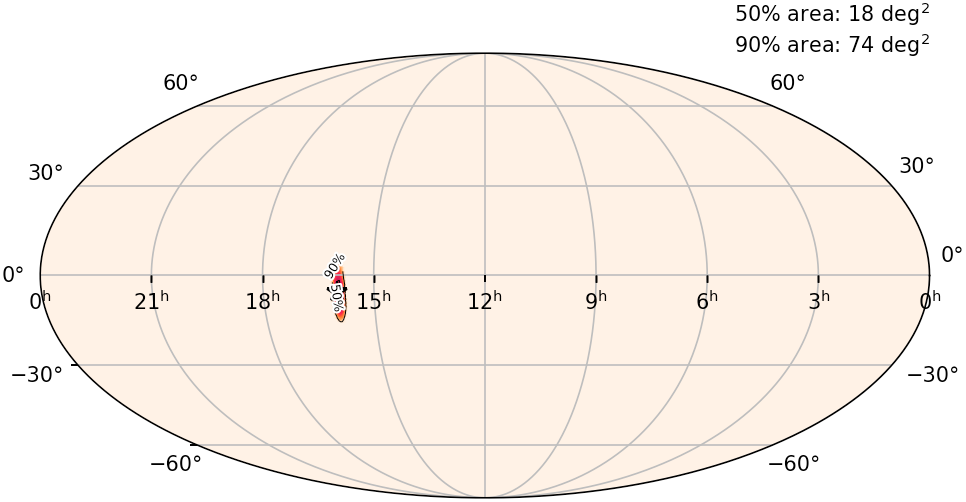}
\includegraphics[width=0.32\textwidth]{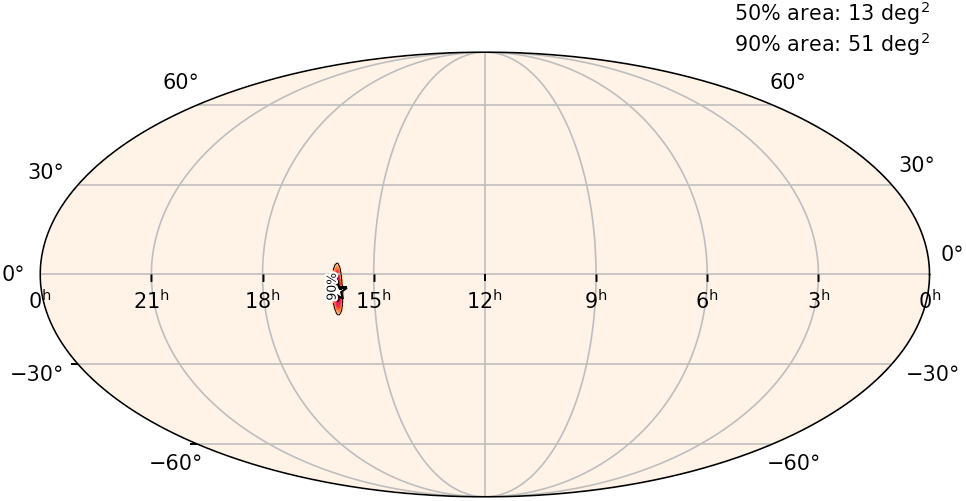}
\includegraphics[width=0.32\textwidth]{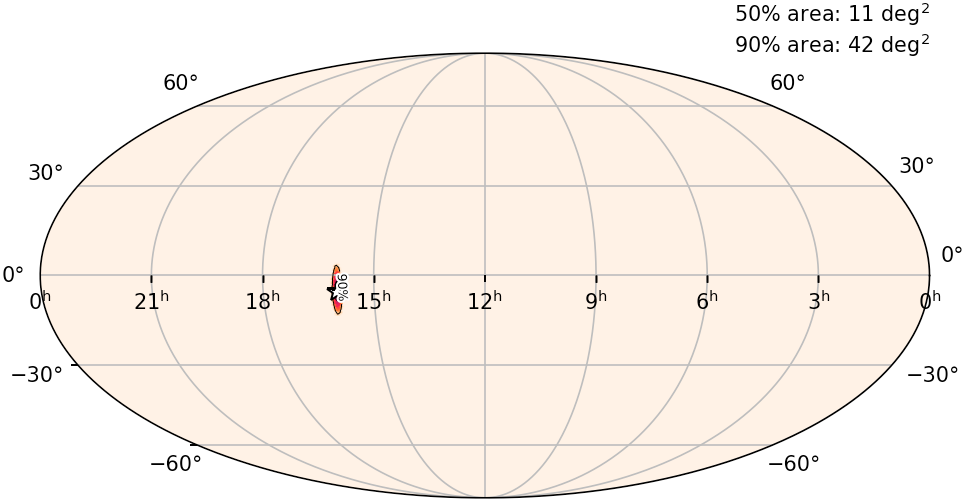}

\vspace{2mm}

\includegraphics[width=0.32\textwidth]{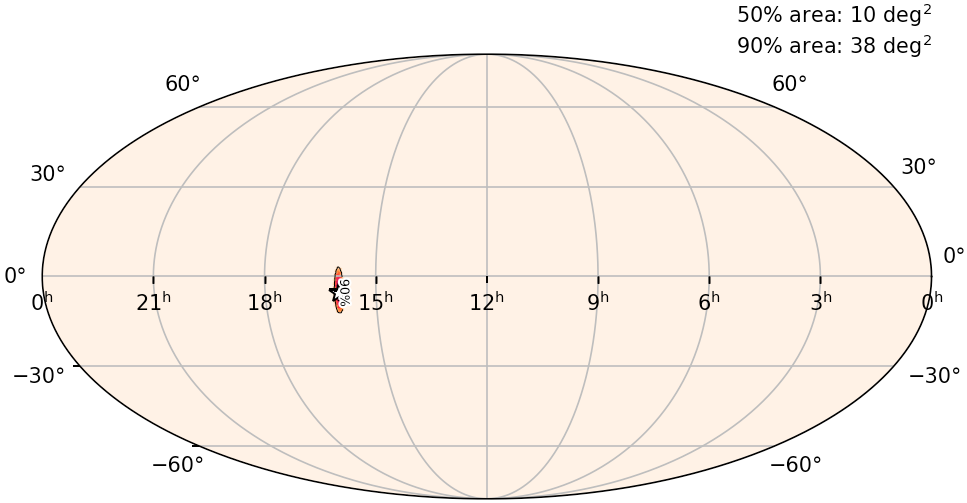}
\includegraphics[width=0.32\textwidth]{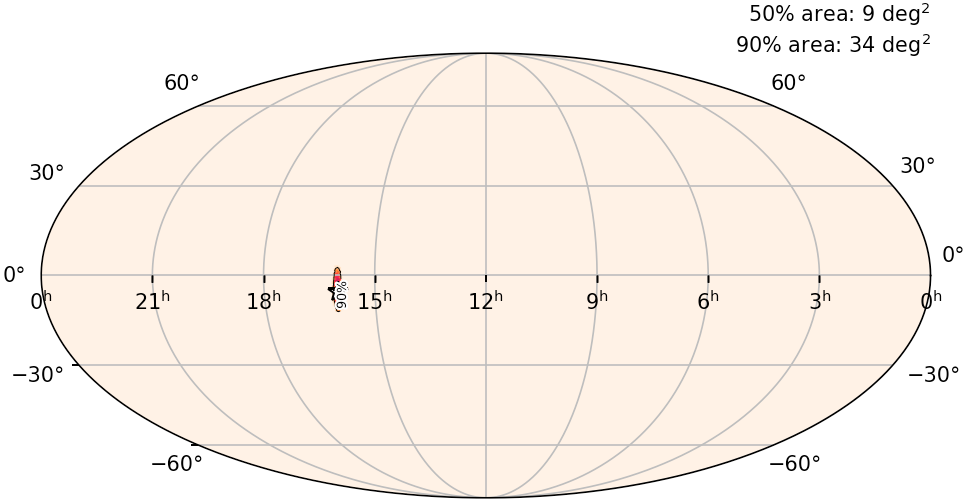}
\includegraphics[width=0.32\textwidth]{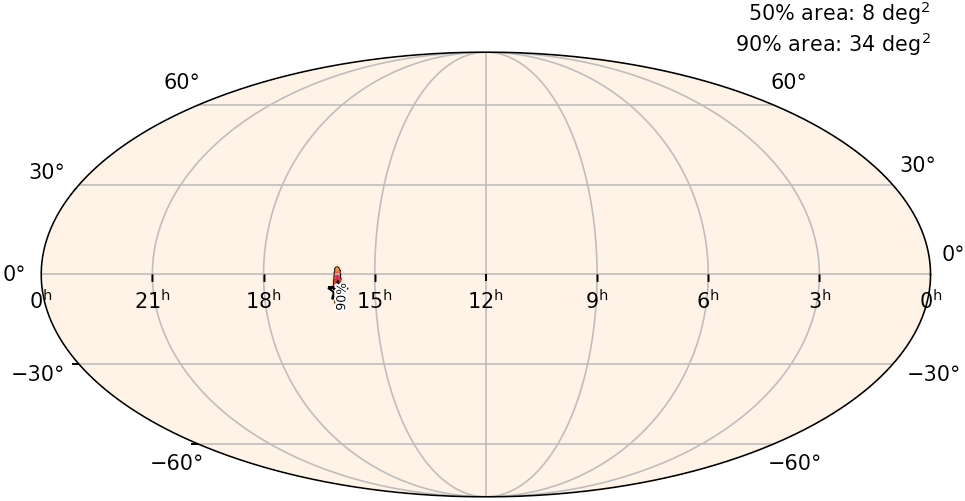}

\vspace{2mm}

\includegraphics[width=0.32\textwidth]{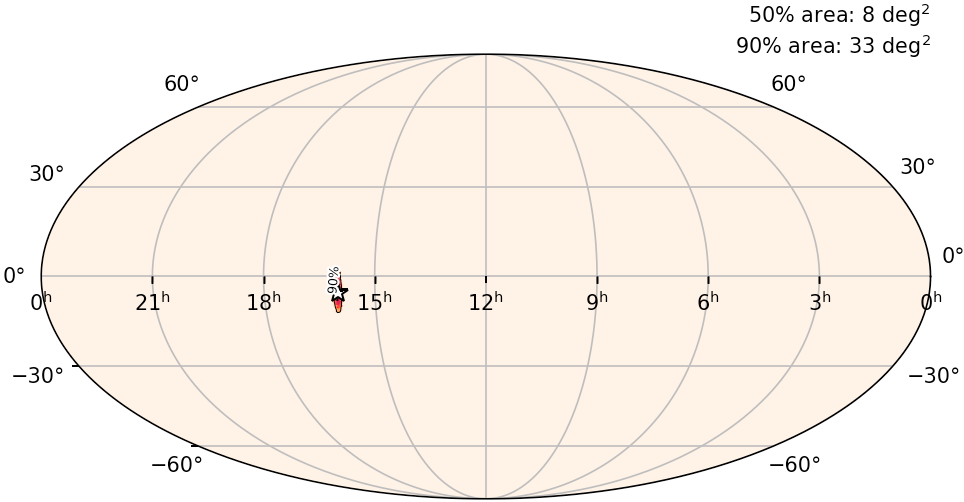}

\caption{
Example sky localization maps for a representative BNS event as a function of KAGRA sensitivity. The panels show increasing KAGRA BNS range from left to right and top to bottom: (a) No KAGRA, (b) 1 Mpc, (c) 2 Mpc, (d) 5 Mpc, (e) 10 Mpc, (f) 20 Mpc, (g) 30 Mpc, (h) 50 Mpc, (i) 75 Mpc, (j) 100 Mpc, (k) 150 Mpc, (l) 200 Mpc, and (m) 250 Mpc. Contours indicate the $50\%$ and $90\%$ credible regions, and the true source location is marked by a star. As KAGRA sensitivity increases, the localization region becomes progressively more compact and less degenerate.
}
\label{fig:skymap_progression}
\end{figure*}

Figure~\ref{fig:skymap_progression} provides an event-level visualization.

At low sensitivity, the localization is dominated by extended ring-like structures. As KAGRA is included, these degeneracies are progressively broken, and the localization collapses into compact regions.

This visual evolution directly illustrates the transition from geometry-dominated localization to high-precision coherent localization.

\section{Conclusion}\label{sec:conclusion}

This study investigates the role of KAGRA in improving gravitational-wave sky localization within the LVK detector network. Using a radiometric, coherence-based localization framework, a systematic injection study of BNS signals quantifies how localization performance depends on detector sensitivity and network configuration.

The results demonstrate that KAGRA improves sky localization through two complementary mechanisms. First, even at low sensitivity, KAGRA provides additional geometric constraints via its independent baseline and antenna response, enabling the network to break degeneracies in sky position and leading to measurable improvements in localization performance~\cite{Grover:2013sha,Fairhurst:2010is,Fairhurst:2009tc,Tsutsui:2020sml}. Second, as KAGRA sensitivity increases, it contributes significant signal-to-noise ratio (SNR) and timing precision, enabling coherent reconstruction across the detector network and further reducing localization uncertainty.

We find that localization performance improves continuously as a function of KAGRA sensitivity, with no sharp threshold separating ``non-contributing'' and ``contributing'' regimes. Importantly, measurable gains are already present at the current KAGRA sensitivity of $\sim 10~\mathrm{Mpc}$. Even in this regime, KAGRA enhances localization through geometric effects, increasing the fraction of well-localized events and reducing the typical localization area.

Additionally, the results identify a BNS range of approximately $30~\mathrm{Mpc}$ as a practical target sensitivity at which KAGRA begins to contribute strongly and systematically to multimessenger gravitational-wave science~\cite{LIGOScientific:2017ync}. At this sensitivity, multiple localization metrics show a clear qualitative improvement: the median localization area approaches $\sim 100~\mathrm{deg}^2$~\cite{Nissanke:2013fka,Magee:2022kkc,Sachdev:2020lfd}, the fraction of well-localized events increases substantially, and the overall distribution of localization areas shifts toward values suitable for efficient electromagnetic follow-up. This benchmark should be interpreted conservatively; rather than marking the onset of usefulness, it represents the point at which KAGRA's contribution becomes robust across the event population.

In addition to improving localization, we find that KAGRA also increases the number of detectable events by expanding the network's sensitivity to lower-SNR signals. At higher sensitivities, this leads to a significant increase in detection rate, complementing the improvements in localization performance. Together, these effects enhance both the quantity and quality of GW observations~\cite{PhysRevLett.116.061102,LIGOScientific:2025yae}.

The event-level sky maps further illustrate the physical origin of these improvements. As KAGRA sensitivity increases, extended ring-like localization structures characteristic of timing degeneracies collapse into compact regions~\cite{Fairhurst:2010is}. This transition reflects the shift from geometry-dominated localization to high-precision coherent localization enabled by a more sensitive and diverse detector network.

Overall, our results highlight that the scientific impact of KAGRA arises not only from its ultimate design sensitivity, but also from its current and near-term performance. Even as a relatively weak detector, KAGRA already contributes meaningful information to the network~\cite{LIGOScientific:2017vwq}. As its sensitivity improves, these contributions become increasingly significant, reinforcing the importance of a geographically distributed and heterogeneous detector network for GW astronomy~\cite{KAGRA:2013rdx,Grover:2013sha}. Our results are generally consistent with similar past studies~\cite{Emma:2024mjs,Pankow:2019oxl}. 

Looking ahead, continued improvements in KAGRA sensitivity will be crucial for enhancing the capabilities of the LVK network, particularly for multimessenger observations of BNS mergers. Our results provide a quantitative framework for understanding these gains and for guiding future detector development to maximize scientific return.

\section{Acknowledgements}
We extend our sincere gratitude to Prof.~Kipp Cannon for his insightful suggestions. 
A.~K.~Y.~L. would like to acknowledge the support from the Croucher Foundation.
O.A.H. acknowledge support by grants from the Research Grants Council of Hong Kong (Project No. CUHK 14304622, 14307923, and 14307724), the start-up grant from the Chinese University of Hong Kong, and the Direct Grant for Research from the Research Committee of The Chinese University of Hong Kong. 

This research has made use of data or software obtained from the Gravitational Wave Open Science Center (gwosc.org)\cite{LIGOScientific:2025snk,KAGRA:2023pio,LIGOScientific:2019lzm}, a service of the LIGO Scientific Collaboration, the Virgo Collaboration, and KAGRA. This material is based upon work supported by NSF's LIGO Laboratory which is a major facility fully funded by the National Science Foundation, as well as the Science and Technology Facilities Council (STFC) of the United Kingdom, the Max-Planck-Society (MPS), and the State of Niedersachsen/Germany for support of the construction of Advanced LIGO and construction and operation of the GEO600 detector. Additional support for Advanced LIGO was provided by the Australian Research Council. Virgo is funded, through the European Gravitational Observatory (EGO), by the French Centre National de Recherche Scientifique (CNRS), the Italian Istituto Nazionale di Fisica Nucleare (INFN) and the Dutch Nikhef, with contributions by institutions from Belgium, Germany, Greece, Hungary, Ireland, Japan, Monaco, Poland, Portugal, Spain. KAGRA is supported by Ministry of Education, Culture, Sports, Science and Technology (MEXT), Japan Society for the Promotion of Science (JSPS) in Japan; National Research Foundation (NRF) and Ministry of Science and ICT (MSIT) in Korea; Academia Sinica (AS) and National Science and Technology Council (NSTC) in Taiwan.

\bibliographystyle{apsrev}
\bibliography{citations}

\end{document}